\documentclass[11pt, a4paper]{article}

\usepackage[T1, T2A]{fontenc}
\usepackage[utf8]{inputenc}
\usepackage[english]{babel}
\usepackage{amsmath, amssymb}
\usepackage[left=2cm, right=2cm, top=2cm, bottom=2cm]{geometry}
\usepackage[displaymath]{lineno}
\usepackage[space]{cite}
\usepackage{xcolor, graphicx}
\usepackage{float}
\usepackage{upgreek}
\usepackage[pdftex,colorlinks=true,linkcolor=cyan,citecolor=red,urlcolor=blue]{hyperref}

\begin{document}
\thispagestyle{empty}

% Контактные данные автора, ответственного за связь с редакцией\\
% Бухарский Николай Дмитриевич\\
% Национальный исследовательский ядерный университет "МИФИ", 115409, г. Москва, Каширское ш., 31\\
% контактный телефон +7 916 935-29-86\\
% e-mail: n.bukharskii@gmail.ru

\newpage
\setcounter{page}{1}
\begin{center}
\noindent {\large\bf Conversion of intense laser pulses into electromagnetic fields with use of extended targets}\\
\vspace{0.6cm}

\noindent N.~Bukharskii$^{1,2}$, Ph.~Korneev$^{1,2}$\\

\vspace{0.3cm}
\small{\noindent $^1$ National Research Nuclear University MEPhI, Moscow, Russia;\\
$^2$ P.~N.~Lebedev Physical Institute, Moscow, Russia\\
e-mail: phkorneev@lebedev.ru}
\end{center}

The effects leading to generation of quasi-stationary and propagating electromagnetic fields during the propagation of a laser-driven current discharge pulse in extended targets are considered. The results of numerical modeling describe the interaction of a relativistically intense ultrashort laser pulse with an extended dense target and its transformation into a powerful discharge current pulse. It is shown that under certain conditions such a pulse can be used to generate strong electromagnetic fields with certain spatio-temporal properties. Particular attention is paid to the influence of the discharge pulse parameters on the properties of the induced electromagnetic fields. Analysis of the discharge current profile in dependence on the target thickness is provided and the dynamics of discharge current pulse profile evolution is studied on the time scale $\sim 10$~ps, when the propagation distance is $\sim 3$~mm.\\

\section*{Introduction}

Modern technologies for generation and amplification of ultrashort laser pulses~\cite{1,2} allow achieving intensities on the level of $10^{18}$~W/cm$^2$ and above~\cite{3}. In such strong laser fields electron motion becomes relativistic~\cite{4}, and the laser-matter interaction physics greatly enriches with new phenomena which may be considered in a wide variety of prospective applications. One of the interesting effects observed in intense laser-matter interaction with extended targets is generation and propagation of discharge current pulses~\cite{5,6}. A discharge current pulse presents a return displacement current caused by the formation of strong positive potential in the interaction region which is induced by escaping of a fraction of fast laser-accelerated electrons into vacuum. The front of such discharge current pulse propagates along the target surface with the velocity close to that of light, according to the experimental observations and numerical modeling~\cite{6}. Duration of the discharge current pulse, its amplitude and its propagation velocity may depend on the laser driver parameters. In Ref.~\cite{6}, the propagation velocity observed to be noticeably lower than that of light, however the values varied from shot to shot. A model explaining the observed slowing down of the discharge current pulse was suggested in the same work based on the collisionless dissipation mechanisms in the propagating discharge wave. At moderate energies of ultrashort laser drivers, it was shown numerically in Ref.~\cite{7} that the duration and amplitude of discharge current pulses are proportional to those of the laser driver, while the propagation velocity remains almost unchanged and very close to the light velocity. With the use of petawatt laser drivers the discharge current amplitude can reach $\sim 10^5$~A, and its duration may, according to the numerical predictions, be reduced to a few tens of femtoseconds provided a femtosecond laser driver is used for its excitation~\cite{7}.

At present, several schemes for generating quasi-stationary magnetic fields with the use of nano-second~\cite{11,25,26} and picosecond~\cite{6,8,9} laser pulses are realized experimentally. For shorter laser pulses only some modeling results are available~\cite{13,27}. The field values that can be obtained with the laser-driven discharge-based schemes vary greatly, indicating that the optimal interaction conditions have not yet been found. In this context, interaction at grazing incidence is of a special interest as it showed relatively high conversion efficiency. For example, numerical modeling performed in this regime for the parameters of a constructed multi-petawatt laser facility XCELS showed that the quasi-static magnetic field values may reach $\sim 10^5$~T~\cite{13}. Magnetic fields created with the considered schemes can be used in laboratory astrophysics, e.g. to create directed plasma flows similar to astrophysical jets~\cite{14}, or to model astrophysically-relevant magnetic reconnection~\cite{10,15}. In addition, strong magnetic field is capable of collimating high-energy charged particle flows~\cite{13,16}, which can be used, e.g. in inertial confinement fusion schemes~\cite{17}.

Depending on the duration of the discharge current pulse, the excited fields may demonstrate transient or quasi-stationary behaviour. At the same time, due to the strong current values, the magnitude of these fields, at least, in the near field region, may reach and even exceed relativistic values for the characteristic wavelength of the excited laser field. Such strong fields are of interest, e.g. for acceleration and manipulation charged particles, generation of intense secondary radiation with controlled properties, studies of fundamental processes in hot magnetized plasma. Determination of the properties of the excited fields in relation to the discharge current pulse parameters is an essential part of such studies.      
\begin{figure}
    \centering
    \includegraphics[width = 8cm]{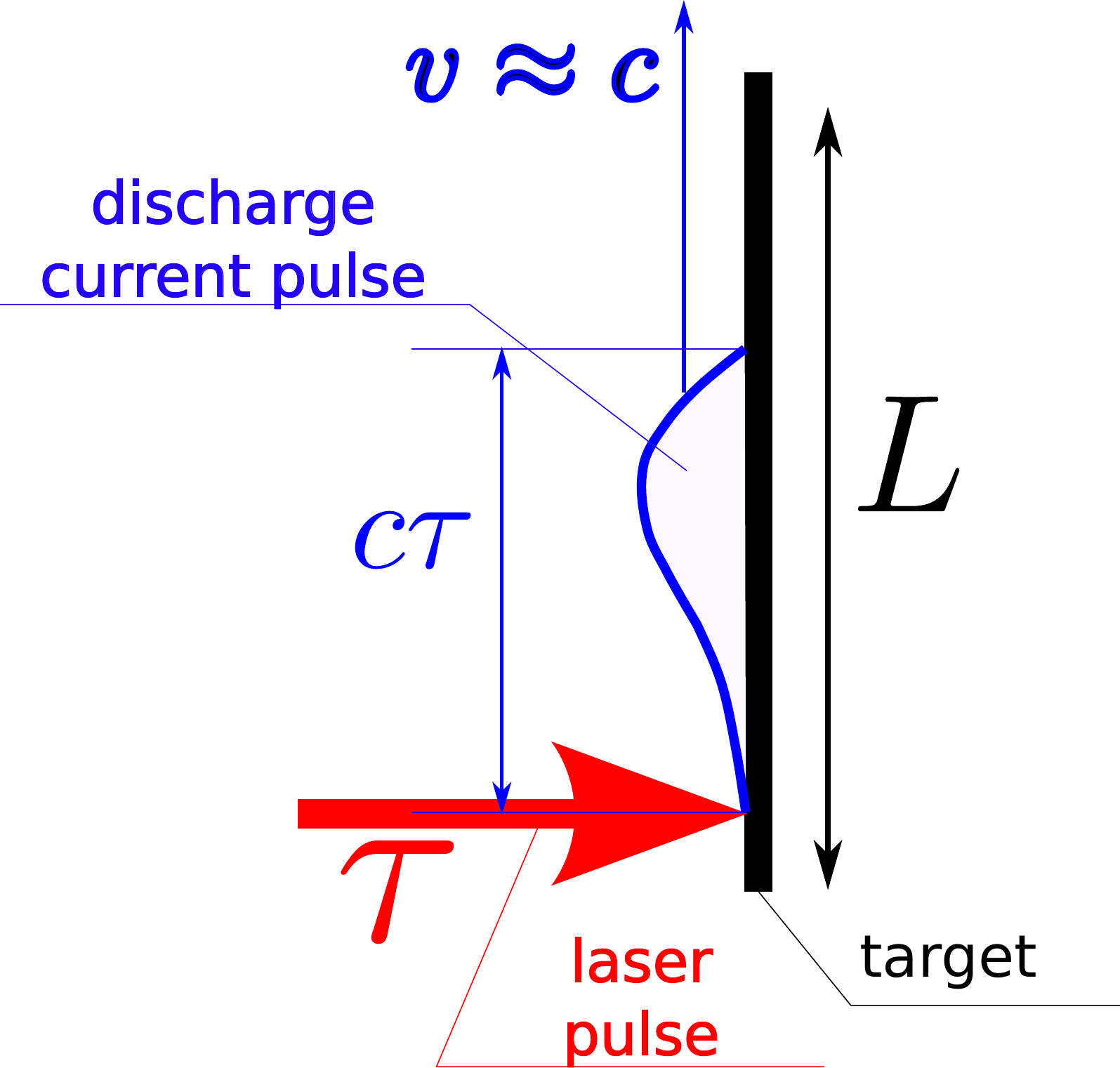}
    \caption{Relation between the target length and the laser driver duration.}
\end{figure}

\section{Dependence of the induced electromagnetic field properties on the discharge current pulse parameters}

One of the most important parameters of interaction is the relation of the laser driver duration to the length of the irradiated target $\xi=c\tau/L$, where $c$ is the speed of light, $\tau$ is the laser driver duration, $L$ is the characteristic length of the irradiated target. In this definition for simplicity it is assumed that the discharge current pulse propagation velocity is equal to that of light, which qualitatively matches the observations, see, e.g. Refs.~\cite{5,6}. For the case $\xi\gg1$, which corresponds to a sufficiently long laser driver, the interaction process is quasi-stationary. In this case, the discharge current pulse transforms to a quasi-stationary discharge current, which essentially means that the displacement current is negligible and both the electric and the magnetic fields in the interaction region are quasi-stationary. In the opposite case of $\xi\ll1$ the interaction is substantially non-stationary.

As it was noted above, one of the potential applications of the discharge currents excited during the laser irradiation of extended targets is generation of strong quasi-stationary magnetic fields. Taking into account the relation between discharge current amplitude and the excited magnetic field magnitude, according to the Ampere-Maxwell law, the conversion efficiency of laser radiation into secondary radiation and fields depends primarily on the absorption efficiency. It in turn depends on the efficiency of the conversion of the irradiating laser beam energy to fast electrons, which form positive potential in the interaction region. With characteristic values of the conversion efficiency of laser beam energy to fast electrons of about $10$~\%, a ratio of a few percent between the energy of the electromagnetic fields associated with the discharge current pulse and the laser driver energy is expected. Estimates show that for the laser pulse energy of $\approx100$~J, and the size of the coil with the excited discharge current of $\approx100$~$\upmu$m, the magnetic field amplitude may reach $\approx1$~kT. This is actually the characteristic value that has been measured in the most successful experiments on optical magnetic field generation~\cite{8,9,10,11,12}.

With the use of nanosecond laser pulses, the relation $\xi\gg1$ holds true, however, when the laser pulse duration is reduced to a few picoseconds and less, and at the same time, the target size is a few hundred microns and more, parameter $\xi$ may become of the order of unity and less. In this case, a compact discharge pulse of the displacement current, which can be interpreted as a polarization surface pulse, is formed in the irradiated conducting target~\cite{30}.
Such localized discharge current may serve as a source of intense secondary radiation provided its spatial scale remains less than the characteristic target scale~\cite{7}. The secondary radiation may be easily generated if such a localized discharge current pulse propagates along a target with a curved surface; for the target sizes of a few hundred microns its spectral maximum corresponds to the terahertz~(THz) frequency range. Sources based on this scheme were studied theoretically~\cite{7,18,19} and showed a high laser-to-THz conversion efficiency and a high output power, up to $\simeq 1$~TW with the use of petawatt laser drivers. At the same time, these studies revealed that a number of key radiation properties, like spatio-temporal profiles, frequency spectra, angular distributions and polarizations, depend on the target surface shape and size. It is expected then, that in the described approach, the THz-radiation properties required for a particular application can be achieved with optimization of the target surface profile. In alternative schemes for generation of powerful THz-radiation pulses with laser-irradiated solid targets, despite similar values of the expected output power with petawatt drivers up to $\simeq 1$~TW~\cite{20}, the control of the aforementioned radiation properties appears to be more limited.

Based on the previous results, see, e.g. the supplementary material for Ref.~\cite{7}, properties of THz-radiation created with the discharge schemes, depend substantially on the parameters of the propagating current pulse. These parameters, in turn, may depend on the target thickness and geometry, they may also change during the discharge current pulse propagation along the surface. The results provided in Ref.~\cite{7} were obtained for a straight plate target with a fixed width of $1$~$\upmu$m which is $\simeq 10$ times less than characteristic values attainable with the current micro-target fabrication technologies. At the same time, evolution of the discharge current pulse in \cite{7} is studied on the time scale of $\simeq 1.5$~ps, which corresponds to the traveling distance of $450$~$\upmu$m. For radiation with a characteristic frequency of $1$~THz this constitutes $1.5$ emission periods. Therefore, it is important to study the discharge current pulse profile evolution on a greater time scale, say $\simeq \left( 5-10 \right)$ emission periods. The reported study aims to address these gaps.

Broadening of the discharge current pulse leads to the change of the parameter $\xi$ which essentially defines the regime of the electromagnetic field generation. Previously, with the use of sub-petawatt laser driver with the laser pulse energy of $50$~J and duration of $0.5$~ps, quasi-stationary magnetic fields  of a few hundred tesla with a lifetime of a few tens of picoseconds were experimentally demonstrated in single-turn coil targets irradiated directly at the open end~\cite{8}. Use of targets irradiated internally under a grazing incidence allowed to reach $\approx 1$~kT~\cite{9,10}. Target sizes in those works were about a few hundred microns which corresponds to $\xi\sim1$. It is the intermediate case of the interaction, which requires some additional analysis to understand the observed quasi-stationary behaviour of the magnetic field in the experiment. If the discharge current pulse broadens significantly, regime of quasi-stationary magnetic field generation corresponds to $\xi\gtrsim1$; in this situation an efficient generation of THz-radiation becomes impossible even for ultrashort laser driver duration. So, localization of the discharge current pulse is the critical parameter defining the physical processes in the considered systems.

\section{Modeling of the discharge current profile}

The process of discharge current pulse formation and propagation along extended targets was studied in this work using numerical modeling. The modeling relied on the kinetic approach based on the Vlasov-Maxwell model well-suited for studies of laser-matter interaction in the relativistic regime when plasma temperatures are high and collisions do not have a significant impact on plasma dynamics. Simulations were performed with the open Particle-In-Cell code Smilei~\cite{21}. The modeling was performed in the two-dimensional (2D) geometry. Such reduction, as was shown in~\cite{22,31}, gives results~(the surface wave and current profiles) which are qualitatively similar to the results obtained with a complete three-dimensional modeling, but at the same time does not require as much computational resources allowing to study different interaction scenarios.

In the frame of the current work, three set of simulations were performed. The first two studied the process of formation and propagation of the discharge current pulses along straight extended targets for different thicknesses of the irradiated plate, $w=5$~$\upmu$m and $w=10$~$\upmu$m. The third set of simulations was dedicated to study of the temporal evolution of the discharge current pulse profile on the time scale of $\simeq 10$~ps which corresponds to the traveling distance of $\simeq 3$~mm. These simulations were performed for the smaller target thickness of $w=5$~$\upmu$m, in order to reduce the amount of plasma in simulation and to lower the computational cost.

In all the simulations, the laser pulse parameters were the same. The laser pulse was introduced into the simulation box by defining an oscillating boundary condition for $B_z$ component on the left boundary. The laser pulse with the wavelength $800$~nm and duration $24$~fs~(FWHM) was focused on the upper boundary of a rectangular plate with a thickness $w$ at an angle of $45^{\circ}$ to the surface. Such irradiation geometry facilitates excitation of the discharge current pulse in the desired direction while the discharge current in the opposite direction appears to be substantially suppressed~\cite{7, 9, 13, 19}. The focal spot diameter was $2$~$\upmu$m~(the radius where the intensity is $1/e^2$ of its maximum value), and the peak intensity was $10^{21}$~W/cm$^2$. Such laser radiation parameters can be obtained at modern petawatt laser facilities, see, e.g. Ref.~\cite{23}.

The target in simulations was defined as an ionized plasma slab consisting of ions with the volume density $n_i = n_c$ and the charge $Z=10$ and of electrons with the volume density $n_e = 10\,n_c$, where $n_c$ is the critical plasma density for the considered laser wavelength of $800$~nm. The chosen density values are a compromise which ensure qualitatively correct description of the interaction process, but at the same time, they are much lower than solid-state densities, which require much higher resolution significantly increase computational resources needed. The spatial resolution~(the cell size) in simulations was $\approx 32$~nm which does not exceed the Debye length for the modeled plasma and allows for resolving its dynamics correctly. The temporal resolution was chosen in accordance with the Courant–Friedrichs–Lewy criterion~\cite{24} as $\approx0.1$~fs. Open boundary conditions were used for the fields and particles. In the third set of simulations, a "moving window" was used, which is realized in Smilei~\cite{21}. There, the simulation domain was periodically shifted along the $x$-axis to follow the discharge current pulse propagating approximately with the light velocity; the size of the simulation domain was $260$~$\upmu$m, and at each step, only a fraction of the whole target matter was modeled. Such approach allowed extending the simulated time duration from $\simeq 1$~ps to $\simeq 10$~ps and obtaining the discharge current pulse profiles for propagating distances of about $3$~mm.

\begin{figure}
    \centering
    \includegraphics[width=16cm]{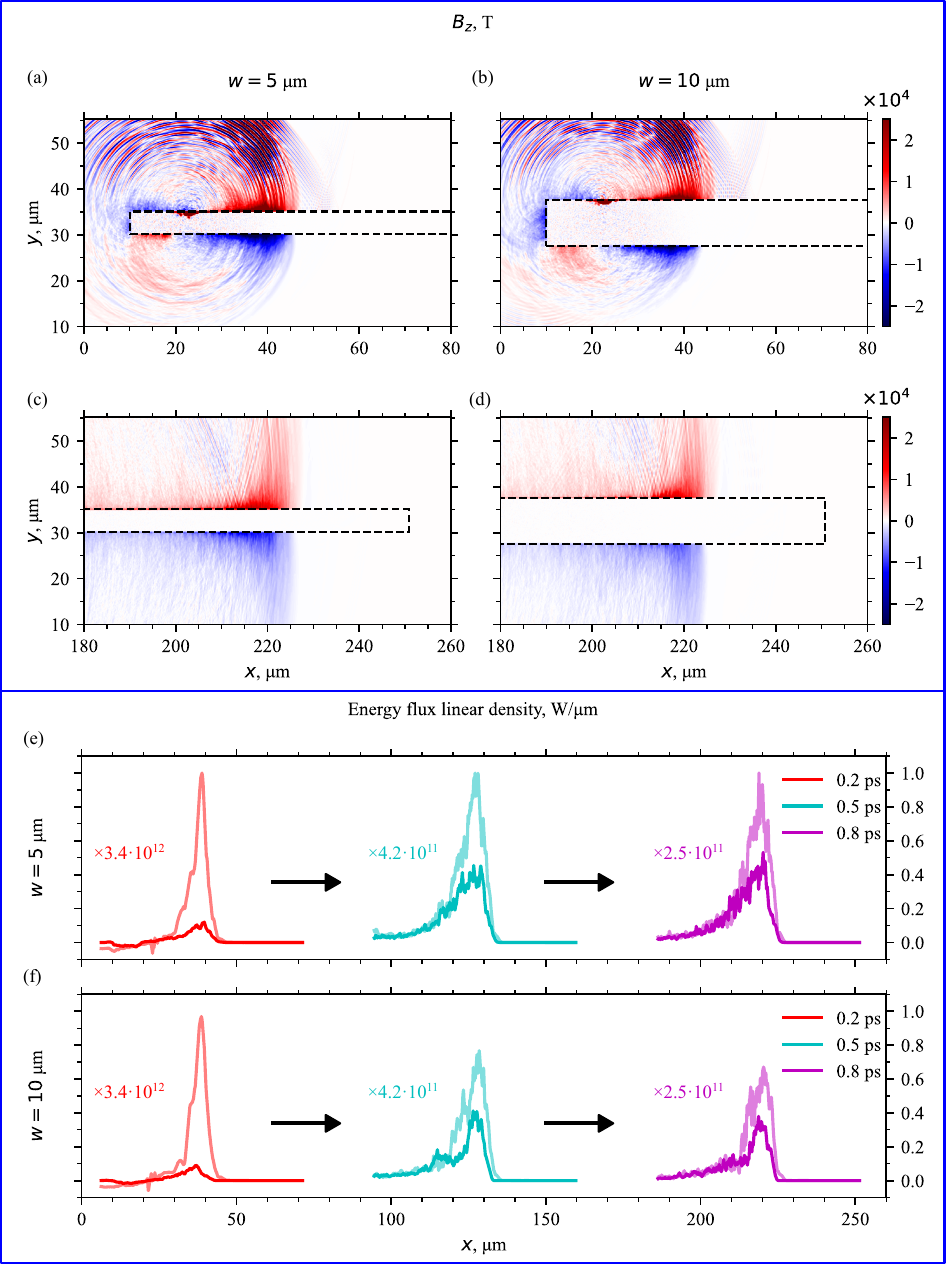}
    \caption{Results of the 2D modeling of the discharge current pulse formation and propagation for straight extended targets with thicknesses of $w=5$~$\upmu$m and $w=10$~$\upmu$m: (a,b)~profiles of the transverse magnetic field component $B_z$ at time moment $t=0.2$~ps; (c,d)~$B_z$ profiles at time moment $t=0.8$~ps; dashed line shows the target surface; (e,f)~profiles of linear energy density carried along the surface for time moments $0.2$~ps, $0.5$~ps and $0.8$~ps, partially transparent curves correspond to the profiles on the upper surface, solid curves -- to the profiles on the lower surface. For ease of comparison, profiles are scaled, the scaling factor is shown near each profile.}
    \label{fig:PIC_1}
\end{figure}
In Fig.~\ref{fig:PIC_1}, the comparison of the modeling results for different target thicknesses, $w=5$~$\upmu$m and $w=10$~$\upmu$m, is presented. Panels (a,b) show profiles of the transverse magnetic field component $B_z$ at early stages of the modeling, at the time moment $t=0.2$~ps. In the presented distributions, one can see the laser pulse being reflected at $45^o$ and the laser-induced surface wave propagating along the surface to the right, with the velocity which appears to be $\left( 2-3 \right)$~\% lower than that of light. The surface wave is induced both on the upper and on the lower surfaces, though the laser induced perturbation reaches the lower surface with a delay. This effect becomes noticeable for a greater plate thickness $w=10$~$\upmu$m. As can be seen in the magnetic field profile in panel (b), the wave propagating along the upper surface somewhat outpaces the wave propagating along the lower one. Later in time, with the increase of the traveling distance, the upper and lower parts of the propagating pulse synchronize: the delay between their front arrival to a given coordinate $x$ decreases and the two waves propagate together along the two opposite plate surfaces. This can be seen in the profiles of the magnetic field component $B_z$ corresponding to time moment $t=0.8$~ps, see panels~(c,d) in Fig.~\ref{fig:PIC_1}. It is worth noting also that in profiles corresponding to the early stages of the interaction, there is also a laser-induced discharge pulse propagating in the opposite direction (left), though its amplitude is a few times lower than that for the main pulse propagating to the right. This is a consequence of the chosen irradiation geometry and the incidence direction of the laser pulse. The surface pulse propagating to the left is partially reflects from the left edge of the plate, contributing to the formation of a long "tail" following the main discharge current pulse, propagating to the right. However, this "tail" is mainly attributed to the currents frozen into surface plasma heated by the main discharge current pulse, so the reflected pulse only slightly modifies it.

Further analysis of the obtained results was carried out using distributions of the linear density of the energy flux carried by the discharge pulse along the surface. This distributions were obtained by numerical integration along the the $y$-axis of the $x$-component of Poynting vector $S = \frac{c}{4\pi}\,\mathbf{E}\times\mathbf{H}$, where $c$ is the light velocity, $\mathbf{E}$ and $\mathbf{H}$ are the electric and magnetic fields in the wave. Linear energy density profiles in the $xy$-plane obtained in such a way are shown in panels~(e,f) of Fig.~\ref{fig:PIC_1} using three curves in each of the panels: the red one corresponds to the profile at the time moment $0.2$~ps, the cyan one corresponds to the profile at the time moment $0.5$~ps and the magenta one corresponds to the profile at the time moment $0.8$~ps. Partially transparent curves in the panels correspond to the profiles on the upper surface of the target, solid curves -- to the profiles on the lower surface. For ease of comparison, the profiles presented in panels~(e,f) are scaled with the scaling factor shown near each profile. The red curves show that initially an order of magnitude more powerful wave is induced on the upper surface which is directly irradiated by the laser pulse, however, subsequently the energy flux carried by the discharge waves redistributes. Later, the profiles show that the energy carried by the wave propagating along the lower surface amounts to $\approx50$~\% of the energy carried by the discharge wave propagating along the upper plate surface. It is worth noting also that a thicker plate, with $w=10$~$\upmu$m is characterized by a greater value of dissipation. According to the results presented in panels~(e,f), one can see that at the early stage ($0.2$~ps), the energy density is almost the same for the plates with $w=5$~$\upmu$m and $w=10$~$\upmu$m, however, for the subsequent time moments $0.5$~ps and $0.8$~ps, the discharge current pulse looses more energy for $w=10$~$\upmu$m. This is probably related to the fact that the plate with $w=10$~$\upmu$m contains more matter and more cold electrons, which may absorb a fraction of the discharge current pulse energy. According to the presented results, the discharge current pulse power decay in case of the thicker plate is $\sim40$~\% for the last time moment of $0.8$~ps. Its profile does not significantly depend on the plate width and remains well-localized on the target scale, so probably can be used for generation of THz-radiation if the target length is $\gtrsim 100$~$\upmu$m. Note that the plate thickness considered in simulations $w=10$~$\upmu$m is very close to the sizes consistent with the modern micro-target fabrication technologies, see, e.g. Ref.~\cite{8}. Therefore, theoretical results obtained in earlier works~\cite{7, 18, 19}, related to the formation of ultrashort discharge current pulses on extended surfaces are expected to hold true for real targets with a thickness of $\gtrsim 10$~$\upmu$m.
\begin{figure}
    \centering
    \includegraphics[width=16cm]{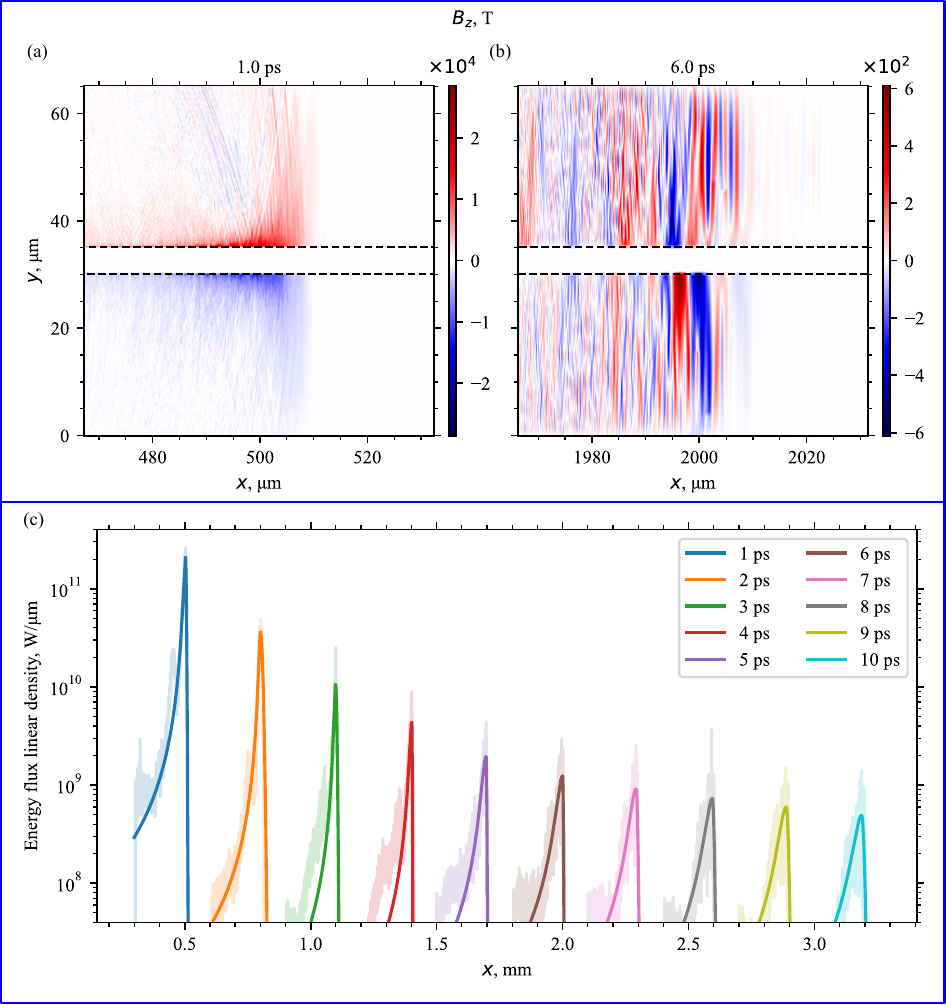}
    \caption{Results of the 2D modeling of the discharge current pulse propagation along the flat plate with the thickness of $w=5$~$\upmu$m and length $\gtrsim 3$~mm: (a,b)~comparison of the profiles of the transverse magnetic field component $B_z$ at time moments $1.0$~ps, $6.0$~ps; dashed line in panels (a-d) shows the target surfaces; (c)~profiles of the linear energy density in $xy$-plane at different time moments, from $1$~ps to $10$~ps, partially transparent curves correspond to the explicit data extracted from the simulation, solid curves correspond to the results of the approximation~(\ref{eq:pulse_shape}).}
    \label{fig:PIC_2}
\end{figure}

For real experiments on the generation of short discharge current pulses on extended surfaces it is of interest to study the temporal dynamics of the discharge current pulse profile evolution for large propagation distances $\gtrsim 3$~mm, on the time scale $\gtrsim 10$~ps which in the context of generation of secondary radiation will correspond to $\gtrsim 10$ emission periods of the radiation with a characteristic frequency $\simeq 1$~THz. Analysis of this temporal dynamics was performed with use of the moving window in the simulations. The obtained results are presented in Fig.~\ref{fig:PIC_2}. In panels (a,b) two-dimensional profiles of the transverse magnetic field component $B_z$ are shown for two different time moments, $t=1.0$~ps~(panel~(a)) and $t=6.0$~ps~(panel~(b)). At the early stage of the propagation, at the time moment $t=1.0$~ps, the discharge current pulse profile presents an unipolar wave with the polarity defined by the direction of the discharge current~\cite{7, 18, 19}. In this case, bulk electrons flow to the left, the discharge current is directed to the right, so $B_z$ component is positive in the upper half-space and negative in the lower half-space. With time, however, the surface wave profile changes, and its unipolar structure modifies, which can be seen in the profile corresponding to time moment $t=6.0$~ps, shown in Fig.~\ref{fig:PIC_2}(b).The wave at this time remains well localized, though its inner structure consists of two different parts with the opposite polarities of the magnetic field, which makes the surface wave more similar to the Sommerfeld-Zenneck wave~\cite{28, 29} with a characteristic duration of one period. Profiles of the linear density of the wave energy flux are shown in Fig.~\ref{fig:PIC_2}(c). Partially transparent curves correspond to the explicit data extracted from the simulation, solid curves -- to the results of the approximation 
\begin{equation}
    f(x) = 
    \begin{cases}
        a\,e^{-\left( \frac{x - x_0}{w_G} \right)^2},\,&x \geq x_0 , \\ 
        a\,\frac{w_L^2}{\left( x-x_0 \right)^2 + w_L^2},\,&x < x_0\, ,
    \end{cases}
    \label{eq:pulse_shape}
\end{equation}
where $a$ is the maximum value of the power density of the discharge current pulse, $x_0$ is the position of the power density maximum on the target surface, $w_G$ and $w_L$ are the widths of the forward and back fronts which can be represented approximately with Gaussian and Lorentzian curves. The chosen dependencies, taking into account the numerical noise in the data extracted from the simulations, provide reasonable description of the discharge wave profile near its maximum, but at the same time does not allow to describe its trailing edge consisting of magnetized plasma. The described shape of the energy density profile remains quite stable and does not demonstrate qualitative changes for discharge current pulse propagation distances of up to $\simeq 3$~mm, despite the fact that the inner structure of the electromagnetic field in the pulse changes as discussed above. At the same time, the peak power density in the pulse, as can be seen in Fig.~\ref{fig:PIC_2}(c), at the distance of $\simeq 3$~mm drops by the two orders of magnitude. 

\begin{figure}
    \centering
    \includegraphics[width=16cm]{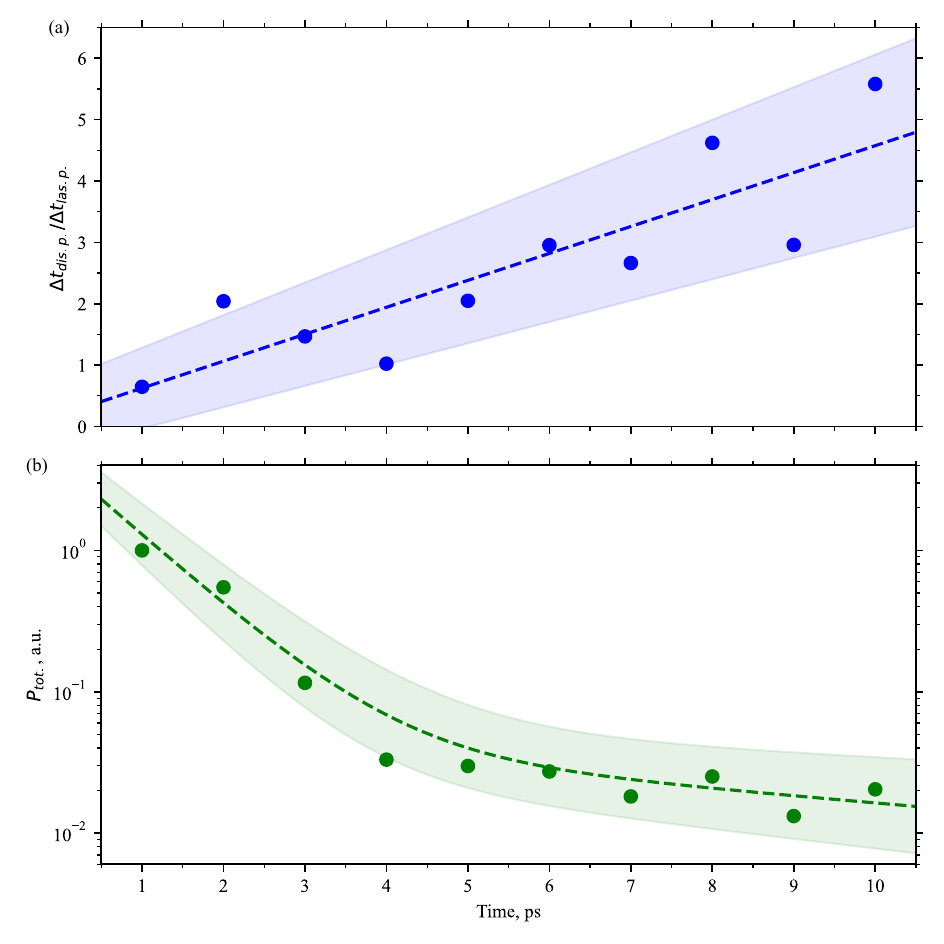}
    \caption{Results of the 2D modeling of the discharge current pulse propagation along a straight extended target with the width $w=5$~$\upmu$m and the length $\gtrsim 3$~mm: (a)~time dependence of the ratio of the discharge current pulse duration $t_{dis.p.}$~(FWHM) to the laser pulse duration $t_{las.p.}$~(FWHM), points show the data extracted explicitly from the simulation, dashed line shows results of an approximation with a linear function; (b)~time dependence of the power emitted by the discharge current pulse, points show the data extracted explicitly from the simulation, dashed line shows results of an approximation with a sum of two decaying exponents. Approximation errors are shown in the plots with half-transparent bands around the curves.}
    \label{fig:PIC_3}
\end{figure}

Analysis of the temporal evolution of the discharge current pulse parameters is given in Fig.~\ref{fig:PIC_3}. 
In panel~(a), the time dependence of the ratio of the discharge current pulse duration $t_{dis.p.}$~(FWHM) to the laser pulse duration $t_{las.p.}$~(FWHM) is presented; the data extracted explicitly from the simulation is shown with points. This dependence is close to linear; an appropriate curve for the linear approximation is shown with a dashed line while half-transparent band shows approximation errors. Thus, discharge current pulse duration grows with traveling distance almost linearly. This trend was not observed in earlier works~\cite{7}, though propagation distances considered there were substantially shorter than $\sim3$~mm as it is here. According to the results presented in this work, the discharge current pulse broadening coefficient amounts to $\approx 1.5$/mm: in $\approx10$~ps which is needed for the discharge current pulse to pass $\approx3$~mm, its width increases by $ 4..5 $ times. Despite such a significant broadening, with the obtained broadening coefficient, the discharge current pulse may still remain well-localized for targets with a characteristic size $\gtrsim 300$~$\upmu$m for propagation distances of $\lesssim 5$~mm, relevant for practical applications. Indeed, if the discharge current pulse is created using the $\simeq 20$~fs laser pulse and has the same duration of $\simeq 20$~fs, its duration at $5$~mm from the point of the initialization would increase to $150$~fs, and it spatial width would increase to $45$~$\upmu$m, which is still an order of magnitude smaller than the assumed target size. Therefore, the possibility for generation of secondary radiation with a characteristic frequency $\simeq 1$~THz still remains realizable. 

An important question is also the power emitted by the discharge current pulse, which would eventually decrease. Results of the analysis of the corresponding temporal dynamics are provided in Fig.~\ref{fig:PIC_3}(b). The obtained dependence takes into account both the obtained dissipation of the discharge current pulse power and its broadening.
Functionally, time dependence of the emitted power can be represented by a sum of two exponents $a_1\,\exp(-t/t_1) + a_2\,\exp(-t/t_2)$ with $t_1 \approx 0.8$~ps and $t_2 \approx 9$~ps. Initially, the emitted power drops rapidly and in the first three picoseconds its value drops by about an order of magnitude. Then, the rate of its decrease decreases, and during the next stage, it falls off on the timescale of $\gtrsim 10$~ps. Therefore, the considered scheme may create the most powerful radiation in the first three picoseconds, after which the discharge current pulse drops substantially, and then reaches quasi-stationary regime, where the emitted power changes slowly, no more than by a few tens of percent in one picosecond. The efficient THz emission therefore may be obtained with targets of $\lesssim 1$~mm length. Increase of the target length would not help to substantially increase the emitted energy. However, the long targets may be considered for THz radiation with lower frequencies. Note also that for certain target geometries, the frequency spectrum can have a strong dependence on the observation direction due to the Doppler effect, which enables generation of radiation in a wide range of frequencies for a fixed target size~\cite{7}.

\section{Conclusion}

In this work, using numerical modeling with the Particle-In-Cell method, we studied the process of conversion of ultrashort relativistically intense laser pulses into discharge current pulses propagating along an extended target surface and capable of creating strong electromagnetic fields and powerful secondary radiation in the THz frequency range. 

The process of femtosecond discharge current pulse formation was studied with the use of extended targets with the spatial sizes (thickness and width) close to real values. Analysis of the obtained results showed that femtosecond discharge current pulses can be formed in plates with thicknesses of $10$~$\upmu$m. They form both on the upper and the lower plate surfaces with almost the same profile shapes. Propagation of the discharge current pulse along a thicker plate is characterized by a greater absorption coefficient, which is probably related to a larger material volume which absorbs the discharge current pulse energy by heating.

In the process of propagation along an extended target with a length $\geq 1$~mm, the width of the discharge current pulse increases almost linearly. However, the rate of this increase is sufficiently low, and for characteristic traveling distance of $\simeq 5$~mm its broadening does not exceed a factor of $10$. It means, in particular, that the discharge current pulse retains its ability to produce secondary THz-radiation.
During the propagation of the discharge current pulse its power drops exponentially, which leads to an exponential decrease of power of the emitted secondary radiation. Calculations show that in $\simeq 3$~ps, the radiated power drops by $\simeq 10$ times.

The results obtained in this work are of a fundamental interest and, besides, may help in preparation of experimental studies on generation of discharge current pulses in extended targets and use of such pulses for the creation of strong quasi-stationary or emitted electromagnetic fields.

\section{Acknowledgements}

The work was supported by the Ministry of science and higher education of the Russian Federation~(Agreement No. 075-15-2021-1361). Authors acknowledge the NRNU MEPhI High-Performance Computing Center and the Joint Supercomputer Center of RAS.


\begin{thebibliography}{99}

\bibitem{1}
Strickland D., Morou G. // Opt. Commun. 1985. V.~56, No.~3. P.~219--221. \href{https://doi.org/10.1016/0030-4018(85)90120-8}{doi: 10.1016/0030-4018(85)90120-8}

\bibitem{2}
Dubietis A., Butkus R., Piskarskas A.\,P. // IEEE Journal of Selected Topics in Quantum Electronics. 2006. V.~12, No.~2. P.~163--172. \href{https://doi.org/10.1109/JSTQE.2006.871962}{doi: 10.1109/JSTQE.2006.871962} 

\bibitem{3}
Yoon J.\,W., Kim Y.\,G., Choi I.\,W., Sung J.\,H., et. al. // Optica. 2021. V.~8. No.~5. P.~630--635. \href{https://doi.org/10.1364/OPTICA.420520}{doi: 10.1364/OPTICA.420520}

\bibitem{4}
Mourou G.\,A., Tajima T, Bulanov S.\,V. // Rev. Mod. Phys. 2006. V.~78. No.~2. no.~309. \href{https://doi.org/10.1103/RevModPhys.78.309}{doi: 10.1103/RevModPhys.78.309}

\bibitem{5}
Quinn K., Wilson P.\,A., Cecchetti C.\,A., Ramakrishna B., et al. // Phys. Rev. Lett. 2009. V.~102. No.~19. no.~194801. \href{https://doi.org/10.1103/PhysRevLett.102.194801}{doi: 10.1103/PhysRevLett.102.194801}

\bibitem{6}
Ehret M., Bailly-Grandvaux M., Korneev P., Apiñaniz J.\,I., et al. // Phys. Plasmas. 2023. V.~30. No.~1. no.~013105. \href{https://doi.org/10.1063/5.0124011}{doi: 10.1063/5.0124011}

\bibitem{7}
Bukharskii N., Korneev Ph. // Matter Radiat. Extremes. 2023. V.~8. No.~4. no.~044401. \href{https://doi.org/10.1063/5.0142083}{doi: 10.1063/5.0142083}

\bibitem{8}
Kochetkov Iu.\,V., Bukharskii N.\,D., Ehret M., Abe Y., et. al. // Sci. Rep. 2022. V.~12. No.~1. no.~13734. \href{https://doi.org/10.1038/s41598-022-17202-2}{doi: 10.1038/s41598-022-17202-2}

\bibitem{9}
Ehret M., Kochetkov Yu., Abe Y., Law~K.\,F\,F., et al. // Phys. Rev. E 2022. V.~106. No.~4. no.~045211. \href{https://doi.org/10.1103/PhysRevE.106.045211}{doi: 10.1103/PhysRevE.106.045211}

\bibitem{10}
Law~K.\,F\,F., Abe Y., Morace A., Arikawa Y., et al. // Phys. Rev. E 2020. V.~102. No.~3. no.~033202. \href{https://doi.org/10.1103/PhysRevE.102.033202}{doi: 10.1103/PhysRevE.102.033202}

\bibitem{11}
Santos J.\,J., Bailly-Grandvaux M., Giuffrida L., Forestier-Colleoni P., et al. // New J. Phys. 2015. V.~17. No.~8. no.~083051. \href{https://doi.org/10.1088/1367-2630/17/8/083051}{doi: 10.1088/1367-2630/17/8/083051}

\bibitem{12}
Law K.\,F.\,F., Bailly-Grandvaux M., Morace A., Sakata S., et al. // Appl. Phys. Lett. 2016. V.~108. No.~9. no.~091104. \href{https://doi.org/10.1063/1.4943078}{doi: 10.1063/1.4943078}

\bibitem{13}
Bukharskii N.\,D., Korneev Ph.\,A. // Bull. Lebedev Phys. Inst. 2023. V.~50. No.~8. P.~869–S877. \href{https://doi.org/10.3103/S1068335623200022}{doi: 10.3103/S1068335623200022}

\bibitem{14}
Albertazzi B., Ciardi A., Nakatsutsumi M., Vinci T., et al. // Science. 2014. V.~346. No.~6207. P.~325–328. \href{https://doi.org/10.1126/science.1259694}{doi: 10.1126/science.1259694}

\bibitem{15}
Palmer C.\,A.\,J., Campbell P.\,T., Ma Y., Antonelli L., et al. // Phys. Plasmas. 2019. V.~26. No.~8. no.~083109. \href{https://doi.org/10.1063/1.5092733}{doi: 10.1063/1.5092733}

\bibitem{16}
Bailly-Grandvaux M., Santos J.\,J., Bellei C., Forestier-Colleoni P., et al. // Nat. Commun. 2018. V.~9, No.~1. no.~102. \href{https://doi.org/10.1038/s41467-017-02641-7}{doi: 10.1038/s41467-017-02641-7}

\bibitem{17}
Sakata S., Lee S., Morita H., Johzaki T., et al. // Nat. Commun. 2018. V.~9, No.~1. no.~3937. \href{https://doi.org/10.1038/s41467-018-06173-6}{doi: 10.1038/s41467-018-06173-6}

\bibitem{18}
Bukharskii N., Kochetkov I., Korneev P. // Appl. Phys. Lett. 2022. V.~120. No.~1. no.~014102. \href{https://doi.org/10.1063/5.0076700}{doi: 10.1063/5.0076700}

\bibitem{19}
Dmitriev E., Bukharskii N., Korneev P. // Photonics. 2023. V.~10. No.~7. no.~803. \href{https://doi.org/10.3390/photonics10070803}{doi: 10.3390/photonics10070803}

\bibitem{20}
Liao G.-Q., Liu H., Scott G.\,G., Zhang Y.-H., et al. // Phys. Rev. X. 2020. V.~10. No.~3. no.~031062. \href{https://doi.org/10.1103/PhysRevX.10.031062}{doi: 10.1103/PhysRevX.10.031062}

\bibitem{21}
Derouillat J., Beck A., Pérez F., Vinci T., et al. // Comput. Phys. Commun. 2018. V.~222. P.~351-373. \href{https://doi.org/10.1016/j.cpc.2017.09.024}{doi: 10.1016/j.cpc.2017.09.024}

\bibitem{22}
Бухарский Н.\,Д., Корнеев Ф.\,А. // Вестник Объединенного института высоких температур. 2022. Т.~8. С.~53–58.

\bibitem{23}
Danson C.\,N., Haefner C., Bromage J., Butcher T., et al. // High Power Laser Sci. Eng. 2019. V.~7. no.~e54. \href{https://doi.org/10.1017/hpl.2019.36}{doi: 10.1017/hpl.2019.36}

\bibitem{24}
Courant R., Friedrichs K., Lewy H. // Math. Ann. 1928. V.~100. P.~32–74. \href{https://doi.org/10.1007/BF01448839}{doi: 10.1007/BF01448839}

\bibitem{25}
Goyon C., Pollock B.B., Turnbull D.P., et.al. // Phys. Rev. E 2017. V. 95, 33208  \href{https://doi.org/10.1103/PhysRevE.95.033208}{doi: 10.1103/PhysRevE.95.033208}

\bibitem{26}
Gao L., Hantao J., Fiksel G., et al. // Physics of Plasmas. 2016. V.~23, No~4. P.~43106 \href{https://doi.org/10.1063/1.4945643}{doi: 10.1063/1.4945643}

\bibitem{27}
Brantov A.V., Korneev Ph., and Bychenkov V.Yu. // Laser Physics Letters. 2019. Vol.~16, no.~6, P.~1 \href{https://doi.org/10.1088/1612-202X/ab1cb4}{doi: 10.1088/1612-202X/ab1cb4}

\bibitem{28}
Zenneck J. Ann. Phys. 1907. Vol.~328. No.~10. P.~846-866. \href{https://doi.org/10.1002/andp.19073281003}{doi: 10.1002/andp.19073281003}

\bibitem{29}
Sommerfeld A. // Ann. Phys. 1909. Vol.~333. No.~4. P.~665-736. \href{https://doi.org/10.1002/andp.19093330402}{doi: 10.1002/andp.19093330402}

\bibitem{30}
Brantov A. V., Kuratov A. S., Aliev Yu. M., et.al. // Physical Review E. 2020. Vol. 102, No. 2, P.~1–5. \href{https://doi.org/10.1103/physreve.102.021202}{doi: 10.1103/physreve.102.021202}

\bibitem{31}
Bukharskii N., Korneev Ph. // Physics of Plasmas, submitted.

\end{thebibliography}
\end{document}